\documentclass[envcountsect]{llncs} 

\usepackage{pstricks}
\usepackage{amsfonts,amssymb}
\usepackage{stmaryrd}
\usepackage{amsmath}
\usepackage{color}
\usepackage{xy}
\xyoption{all}
\CompileMatrices

\mathcode`\<="4268 
\mathcode`\>="5269 
\mathchardef\gt="313E 
\mathchardef\lt="313C 

\newcommand{\starr}{\star}

\newcommand{\id}{{\rm id}}

\newcommand{\mnd}{{\scriptstyle \nabla}}
\newcommand{\unt}{{\scriptstyle \bot}}
\newcommand{\cmn}{{\scriptstyle \Delta}}
\newcommand{\cun}{{\scriptstyle \top}}
\newcommand{\rox}{\owedge}
\newcommand{\orez}{\Ydown}
\newcommand{\repr}{\Upsilon}

\newcommand{\FHilb}{{\sf FHilb}}
\newcommand{\FVec}{{\sf FVec}}
\newcommand{\Rel}{{\sf Rel}}
\newcommand{\Set}{{\sf Set}}
\newcommand{\FRel}{{\sf FRel}}
\newcommand{\FSet}{{\sf FSet}}

\renewcommand{\to}{\xymatrix@C-.5pc{\ar[r]&}}
\newcommand{\ot}{\xymatrix@C-.5pc{& \ar[l]}}
\newcommand{\tto}[1]{\xymatrix@C-.5pc{\ar[r]^{#1}&}}
\newcommand{\oot}[1]{\xymatrix@C-.5pc{&\ar[l]_{#1}}}
\newcommand{\mono}{\xymatrix@C-.5pc{\ar@{>->}[r]&}} 
\newcommand{\epi}{\xymatrix@C-.5pc{\ar@{->>}[r]&}}
\newcommand{\mmono}[1]{\xymatrix@C-.5pc{\ar@{>->}[r]^{#1}&}} 
\newcommand{\eepi}[1]{\xymatrix@C-.5pc{\ar@{->>}[r]^{#1}&}}
\renewcommand{\mapsto}{\xymatrix@C-.5pc{\ar@{|->}[r]&}}
\newcommand{\mmapsto}[1]{\xymatrix@C-.5pc{\ar@{|->}[r]^{#1}&}}
\newcommand{\inclusion}{\xymatrix@C-.5pc{\ar@{^{(}->}[r] &}}
\newcommand{\iinclusion}[1]{\xymatrix@C-.5pc{\ar@{^{(}->}[r]^{#1}&}}
\newcommand{\rrel}[1]{\xymatrix@C-.5pc{\ar[r]|{|}^{#1}&}}
\newcommand{\rel}{\xymatrix@C-.5pc{\ar[r]|{|}&}}
\newcommand{\ler}{\xymatrix@C-.5pc{&\ar[l]|{|}}}
\newcommand{\ller}[1]{\xymatrix@C-.5pc{&\ar[l]|{|}_{#1}}}

\newcommand{\CCC}{{\cal C}}

\newcommand{{\MMM}}{{\cal M}}

\newcommand{{\PPP}}{{\cal P}}

\renewcommand{\Bbb}{\mathbb}

\newcommand{\DDd}{{\Bbb D}}

\newcommand{\ZZz}{{\Bbb Z}}

 \def\pushright#1{{
    \parfillskip=0pt            
    \widowpenalty=10000         
    \displaywidowpenalty=10000  
    \finalhyphendemerits=0      
    \leavevmode                 
    \unskip                     
    \nobreak                    
    \hfil                       
    \penalty50                  
    \hskip.2em                  
    \null                       
    \hfill                      
    {#1}                        
    \par}}                      

 \def\qed{\pushright{$\square$}\penalty-700 \smallskip}

\newenvironment{prf}[1]{\begin{trivlist} \item[{\bf ~Proof}#1.]}%
{\qed\end{trivlist}}

\newcommand{\be}[1]{\begin{#1}}

\newcommand{\ee}[1]{\end{#1}}
\newcommand{\beq}{\begin{equation}}
\newcommand{\eeq}{\end{equation}}
\newcommand{\ba}[1]{\begin{array}{#1}}
\newcommand{\ea}{\end{array}}
\newcommand{\bea}{\begin{eqnarray}}
\newcommand{\eea}{\end{eqnarray}}
\newcommand{\bear}{\begin{eqnarray*}}
\newcommand{\eear}{\end{eqnarray*}}
\newcommand{\bpr}{\begin{prf}{}}
\newcommand{\epr}{\end{prf}}
\newcommand{\bprf}[1]{\begin{prf}{#1}}
\newcommand{\eprf}{\end{prf}}

\newtheorem{thm}{Theorem}[section]
\newtheorem{defn}[thm]{Definition}
\newtheorem{prop}[thm]{Proposition}
\newtheorem{cond}{}[thm]

\begin{document}
\title{Quantum and classical structures\\ in nondeterminstic computation}

\author{Dusko Pavlovic\thanks{Supported by ONR and EPSRC.}\\%
\institute{Kestrel Institute and
Oxford University}
\email{\small Email:~dusko\char64\{kestrel.edu,comlab.ox.ac.uk\}%
}}
\date{}
\maketitle

\begin{abstract}
In categorical quantum mechanics, classical structures characterize the classical interfaces of quantum resources  on one hand, while on the other hand giving rise to some quantum phenomena. In the standard Hilbert space model of quantum theories, classical structures over a space correspond to its orthonormal bases. In the present paper, we show that classical structures in the category of relations correspond to direct sums of abelian groups.  Although relations are, of course, not an interesting model of quantum computation, this result has some interesting computational interpretations. If relations are viewed as denotations of {\em nondeterministic\/} programs, it uncovers a wide variety of non-standard quantum structures in this familiar area of classical computation. Ironically, it also opens up a version of what in philosophy of quantum mechanics would be called an ontic-epistemic gap, as it provides no interface to these nonstandard quantum structures.
\end{abstract}

\section{Introduction}
Classical structures came to be a useful algebraic tool for analyzing the conceptual foundations of quantum computation \cite{PavlovicD:QMWS,Coecke-Duncan,PavlovicD:CQStruct}. They characterize the classical interfaces of quantum resources on one hand, and generate entanglement structures, and other essentially quantum phenomena on the other. In the standard, Hilbert space model of quantum theories, classical structures over a space exactly correspond to its orthonormal bases. In nonstandard models, however, they provide a generic conduit to the classical and the quantum features.

Categorical quantum mechanics, initiated in \cite{Abramsky-Coecke}, axiomatizes some basic quantum phenomena in the framework of dagger-compact categories. This remarkably rich yet succinct structure has arisen in part from the experience gathered in semantics of programming languages. The most direct source are probably Abramsky's {\em interaction categories\/} \cite{AbramskyS:Interaction,PavlovicD:SIC}, developed to capture the idea of concurrent programs as {\em relations extended in time}. As a consequence, categories of relations,  in all their various flavors arising from various resources \cite{BarrM:exaccs,Carboni-Walters,Freyd-Scedrov,PavlovicD:mapsI,PavlovicD:mapsII}, provide models of categorical quantum mechanics, albeit degenerate because of the trivial dagger structure. Nevertheless, the notion of a classical structure  over relations is well defined. In the present paper, we provide a complete characterization of classical structures over relations.\footnote{Formally, we work with relations within a given universe of sets. Each of the relational formalisms proposed in the above references will suffice for this.} 

But what is the relevance and meaning of such a result? Although some relationally based "toy models" of certain quantum phenomena \cite{SpekkensR:toy} have awaken a lot of interest, a category of relations itself is a rather degenerate model of quantum computation. Its duality and scalar structures in particular seem too simple to accommodate the complex interactions between the quantum  and the classical phenomena. --- It is therefore only more surprising that, even in this simple framework, classical structures seem to have an interesting story to tell.

\subsection*{Outline of the paper} In section \ref{Algebras}, we summarize the definitions of classical and quantum structures, recall their basic properties, and describe the standard, and some nonstandard examples. In section \ref{Simple}, we describe a rich source of nonstandard examples of classical structures in the category $\Rel$ of relations: every Abelian group gives a nonstandard classical structure. In fact, these are exactly the indecomposable classical structures. In section \ref{All}, we show that every classical structure in $\Rel$ must arise as a direct product of indecomposables. This provides a complete characterisation of classical structures in $\Rel$. In section \ref{Conclusions}, we summarize the meaning of this characterization, and discuss the questions that it raises.

\subsubsection*{Notation.} To describe relations on finite sets, we often find it convenient to use von Neumann's representation of ordinals, where $0=\emptyset$ is the empty set, and $n = \{0,1,\ldots, n-1\}$. Moreover, the pairs $<i,j>\in n\times n$ are often abbreviated to $ij\in n\times n$.

\section{Algebras for abstraction and duality}\label{Algebras}
We begin by introducing classical structures as the classical interface of quantum resources. To justify their algebraic content, we delve into a conceptual reconstruction of their role. A reader only interested in their structure should skip the next subsection. 

\subsection{Program abstraction and quantum computation}
Abstraction is the essence of programming. The first example of program abstraction are probably G\"odel's numberings of primitive recursive functions \cite{GodelK:incompleteness}. G\"odel's construction demonstrated that recursive programs, specifying entire families of computations (of the values of a function for all its inputs), can be stored as data. Von Neumann later explicated this as the fundamental principle of computer architecture. Kleene, on the other side, refined the idea of program abstraction into the fundamental lemma of recursion theory: the s-m-n theorem \cite{KleeneS:smn}. Church, finally\footnote{Although Church's paper appeared three years earlier than Kleene's, Church's proposal is the final step in the conceptual development of function abstraction as the foundation of computation.} proposed the formal operations of variable abstraction and data application as the driving force of all computation \cite{ChurchA:1940}. This proposal became the foundation of functional programming.

But what is variable abstraction? What is a variable? We use them so often that it is sometimes hard to tell. A variable is adjoined to a ring, as an indeterminate element, to generate the ring of polynomials. A programmer denotes by a variable a piece of data that will only be determined when the program is run. The variable captures all the possible values of this piece of data that may arise at run time. The operation of abstraction of a variable binds all of its occurrences (within the declared scope) to the interface where its values will be read, when given. The operation of application substitutes these values for the variable.

A variable is thus a tool for propagating as-yet-unknown data through a program (or through an algebraic structure, etc). The crucial capability of such a tool is that it allows the data to be {\em copied\/} wherever they are needed, or {\em deleted\/} where they are not needed. While the classical computation, as the above early references show, was built upon this capability as its very foundation --- it is a fundamental property of quantum data that they generally cannot be copied or deleted \cite{Wooters:NoCloning,Dieks:NoCloning,BraunsteinS:NoDeleting,AbramskyS:NoCloning}.

A classical structure formalizes this distinction: its first feature is a comonoid $X\otimes X\oot{\cmn} X \tto{\cun} I$, where $\cmn$ can be used to copy and $\cun$ to delete a piece of data. A datum $I\tto{\psi} X$ is classical if it can be copied, in the sense that $\cmn\psi = \psi\otimes \psi$, and deleted, in the sense that $\cun\psi = \id_I$. This turns out to be exactly what is needed to support variable abstraction in monoidal categories. In the framework of dagger-monoidal categories, the requirement that abstraction preserves the dagger induces the rest of classical structure \cite{PavlovicD:MSCS97,PavlovicD:Qabs}.

\subsection{Frobenius algebras}\label{Frobenius}
\paragraph{Framework.} Let $(\CCC,\otimes,I)$ be a monoidal category. With no loss of generality, we assume that the tensor is strictly associative and unitary, i.e. that the objects of $\CCC$ form an ordinary monoid with respect to $\otimes$ and $I$. Every monoidal category is equivalent to one which is strict in this sense. We call the arrows from $I$ {\em vectors}, and write $\CCC(X) = \CCC(I,X)$. 

\begin{defn}\label{frobdef}
The structure of a {\em Frobenius algebra\/} $X$ in $\CCC$ consists of
\begin{itemize}
\item 
an internal monoid $X\otimes X\tto{\mnd} X \oot{\unt} I$, and
\item 
an internal comonoid $X\otimes X\oot{\cmn} X \tto{\cun} I$,
\end{itemize}
such that the following diagram commutes
\begin{equation}\tag{{\sf fro}}
\begin{split}
\xymatrix{
X\otimes X \ar[dr]^-{\mnd} \ar[dd]_-{\cmn\otimes X} \ar[rr]^{X\otimes \cmn}&& X\otimes X\otimes X \ar[dd]^-{\mnd\otimes X} \\
& X \ar[dr]^-{\cmn} \\
X\otimes X \otimes X \ar[rr]_-{X\otimes \mnd} && X\otimes X
}
\end{split}
\eeq
A Frobenius algebra $(X,\mnd,\cmn,\unt,\cun)$ is {\em special\/} if its monoid and comonoid structures are normalized, in the sense that the diagram
\beq\tag{{\sf spe}}
\begin{split}
\xymatrix{
& X\otimes X \ar[dr]^-{\mnd} \\
X \ar[rr]_{\id} \ar[ur]^-{\cmn}&& X}
\end{split}
\eeq
also commutes.
\end{defn}

\begin{prop}\label{unique}
In every monoidal category, being a special Frobenius algebra is a  property of the monoid  $(X,\mnd,\unt)$. More precisely, if both $(X,\mnd,\cmn_1,\unt,\cun_1)$ and $(X,\mnd,\cmn_2,\unt,\cun_2)$ are special Frobenius algebras, then $\cmn_1 = \cmn_2$ and $\cun_1 =\cun_2$.

Dually, being a special Frobenius algebra can be viewed as a property of the comonoid $(X,\cmn,\cun)$.

The monoid part $(X,\mnd,\unt)$ of a special Frobenius algebra is abelian if and only if the corresponding comonoid part $(X,\cmn,\cun)$ is.
\end{prop}

Much of the power of the Frobenius algebra structure arises from the way in which it gives rise to dualities.

\subsubsection{Duality.} A {\em duality\/} in a monoidal category $\CCC$ consists of two objects and two arrows, written $(\eta,\varepsilon): X\dashv X^\ast$, where 
\begin{itemize}
\item the copairing $I\tto{\eta}X^\ast \otimes X$ and 
\item the pairing $X\otimes X^\ast \tto{\varepsilon}I$ 
\end{itemize}
are required to satisfy the equations 
\[ (\varepsilon\otimes X)(X\otimes \eta) = X\qquad \qquad(X^\ast\otimes \varepsilon)(\eta\otimes X^\ast) = X^\ast\]
If every object $X\in \CCC$ has a chosen dual $X^\ast$, then the duality can be extended to a functor $(-)^\ast:\CCC^{op}\to \CCC$, which maps $A\tto{f}B$ to 
\[f^\ast: B^\ast\tto{\eta B^\ast} A^\ast AB^\ast \tto{AfB^\ast} A^\ast BB^\ast \tto{A^\ast \varepsilon} A^\ast\] 

\subsubsection{Frobenius algebras and dualities.} Every Frobenius algebra $X$ induces
\begin{itemize}
\item a pairing $\varepsilon: X\otimes X\tto{\mnd} X \tto{\cun} I$, and
\item a copairing $\eta: I\tto{\unt}X\tto{\cmn} X\otimes X$,
\end{itemize}
which together make $X$ into a self-dual object, with $X^\ast = X$. Categorically, this means that $X$ is adjoint to itself at the same time on the left and on the right, if the monoidal category $\CCC$ is viewed as a bicategory with a single 0-cell. E.g., if this 0-cell is a category $\DDd$ and  $\CCC = \DDd^\DDd$ is the category of endofunctors $F,G\ldots:\DDd \to \DDd$, with the natural transformations as the arrows between them, and with the composition playing the role of the tensor $F\otimes G = F\circ G$, then 
\begin{itemize}
\item the monoid $F\otimes F\tto{\mnd} F \oot{\unt} Id$ makes the functor $F$ into a monad,
\item the comonoid $F\otimes F\oot{\cmn} F \tto{\cun} Id$ makes it into a comonad, and
\item condition {\sf (fro)} makes the pairing $\varepsilon: FF\to Id$ and copairing $\eta: Id \to F$ into an adjunction $F\dashv F$.
\end{itemize}
This functorial setting was first described by Lawvere \cite{Lawvere:Ordinal}, who characterized it by requiring the commutativity of the following diagrams of natural transformations
\[
\xymatrix{
FFF \ar[d]_-{\mnd F} & FF \ar[l]_-{F\cmn} \ar[r]^-{\cmn F} \ar[d]^-{\mnd} & FFF \ar[d]^-{F\mnd}  & FF \ar[d]_-{\cmn F} & F \ar[l]_-{\unt F} \ar[r]^{F\unt } \ar[d]^-{\cmn}& FF\ar[d]^-{F\cmn} \\
FF \ar[r]_-{\cun F} & F & FF\ar[l]^-{F\cun }  & FFF \ar[r]^-{F\mnd} & FF & FFF \ar[l]_-{\mnd F}
}
\]
and attached the name of Frobenius to such structures. The equivalent but simpler condition {\sf (fro)} first appeared in Carboni and Walters' work \cite{Carboni-Walters}, characterizing relations as a cartesian bicategory. The geometric meaning of {\sf (fro)} in  the category of cobordisms brought the same condition to prominence in the categorical version of Topological Quantum Field Theory \cite{KockJ:book}. Finally, its role in supporting a generic form of abstraction, on which the interface between the classical and the quantum computation turns out to be based 
\cite{PavlovicD:Qabs}, made it into an important piece in categorical Quantum Mechanics \cite{Abramsky-Coecke,PavlovicD:QMWS}.

\subsection{Classical and quantum structures}\label{Quantum}
\paragraph{Framework.} Categorical quantum mechanics actually requires a slight extension of monoidal categories: besides the monoidal structure, the category $\CCC$ should come equipped with a functor $(-)^\ddag : \CCC^{op}\to \CCC$, which is identity on the objects, and involutive on the arrows, i.e. satisfying $f^{\ddag\ddag} = f$. The arrows $u$ such that $u^\ddag u = \id$ and $uu^\ddag = \id$ are called unitary. All monoidal coherences in a dagger  monoidal category are required to be unitary. In the strict case, this boils down to the requirement that the symmetries are unitary, since the other coherences are already identities. 

The abstract structure of a  {\em symmetric dagger-monoidal\/} category  $(\CCC,\otimes, I, \ddag)$ turns out to support the main constructions of quantum mechanics, normally presented using Hilbert spaces \cite{Abramsky-Coecke,SelingerP:CP,PavlovicD:QMWS}. 

\begin{defn}
A special Frobenius algebra $(X,\mnd,\cmn,\unt,\cun)$ in a symmetric dagger-monoidal category  $\CCC$ is called a  {\em classical structure} if its monoid and its comonoid parts are
\begin{itemize}
\item adjoint, i.e. $\mnd = \cmn^\ddag$ and $\unt = \cun^\ddag$
\item abelian, i.e. $\mnd\circ (a\otimes b) = \mnd\circ (b\otimes a)$.
\end{itemize}
\end{defn}

\subsubsection*{Interpretation.} In categorical quantum mechanics, classical structures can be used to distinguish the classical resources from the quantum resources. On one hand, each classical structure extracts the {\em classical elements}. On the other hand, it supports the entanglement phenomena, implemented through {\em quantum structures}. We now recall these concepts from \cite{PavlovicD:QMWS}.

\be{defn}
A {\em quantum structure\/} in a dagger-monoidal category is a pair $(X,\eta)$, such that $\eta:I\to X\otimes X$ and $\eta^\ddag:X\otimes X \to I$ make $X$ self-dual, i.e. $(\eta,\eta^\ddag): X\dashv X$.
\end{defn}

\begin{prop}
Every classical structure induces a quantum structure, with the pairing $\varepsilon: X\otimes X\tto{\mnd} X \tto{\cun} I$, and the copairing $\eta: I\tto{\unt}X\tto{\cmn} X\otimes X$.
\end{prop}

Several classical structures may induce the same quantum structure. Some quantum structures do not arise from a classical structure.

\begin{definition} {\em Classical elements\/}\footnote{In the Hopf algebra theory, the elements that satisfy the same conditions are called {\em set-like}.} for a classical structure $X$ in $\CCC$ are the arrows $\varphi\in \CCC(X)$ such that $\cmn \varphi = \varphi \otimes \varphi$ and $\cun \varphi = \id_I$.
\end{definition}

Classical elements are thus just those vectors that can be copied and deleted. On the other hand, the entanglement capability of quantum structures is obtained by applying the copying facility of a classical structure to non-classical elements, such as the monoid unit of the classical structure itself.

\subsection{Examples}\label{Examples1}
The trivial example of a classical structure, present in every monoidal category, is the tensor unit $I$: the canonical isomorphisms $I\otimes I \cong I$ make it into a special Frobenius algebra. In the categories $(\sf FVec,\otimes,K)$ of finitely dimensional vector spaces and $(\sf FHilb,\otimes,K)$ over any field $K$, a choice of base $|0>,|1>,\ldots |n>\in X$ makes each space $X$ into a classical structure, defined by the linear operators
\begin{align*}
\cmn |i>  &=   |ii>& \cun |i>  &=  1
\end{align*}
In the monoidal category $(\Rel,\times, 1)$, every object comes with a similar classical structure
\begin{align*}
\cmn (i)  & =   \left\{ii\right\}&\cun (i) & =  0
\end{align*}
where $ij$ abbreviates the pair $<i,j>\in X\times X$, and $0$ is the unique element of $1$.  Both of these families of classical structures are induced by the cartesian structure of the category $\FSet$ of finite sets, canonically embedded in $\FVec$ and $\FRel$. We call them {\em standard\/} classical structures. They are characterized and analyzed in detail in \cite{Carboni-Walters}. In \cite{PavlovicD:MSCS08}, it has been shown that all classical structures in $\FHilb$ are standard. Very recently \cite{Coecke-Edwards}, though, Bill Edwards and Bob Coecke noticed a nonstandard classical structure $2\times 2 \ller{\rox} 2 \rrel{\orez} 1$ over the set $2 = \{0,1\}$, defined as follows
\be{align*}
\rox(0) & =  \{00,11\} & \orez(0) & =  \{0\}\\
\rox(1) & =  \{01,10\}&\orez(1) & =  0
\end{align*}
Note that both the standard classical structure $2\times 2 \ller{\cmn} 2 \rrel{\cun} 1$ and the nonstandard classical structure $2\times 2 \ller{\rox} 2 \rrel{\orez} 1$ induce the same quantum structure $1 \rrel{\eta} 2\times 2$, relating $0$ with $00$ and $11$.
But this turns out to be an exception. E.g., a little trial and error leads to the following nonstandard classical structure $3\times 3 \ller{\rox} 3 \rrel{\orez} 1$, where $3 = \{0,1,2\}$
\begin{align*}
\rox(0) & =  \{00,12,21\} & \orez(0) & =  \{0\}\\
\rox(1) & =  \{22,01,10\} &\orez(1) & =0\\
\rox(2) & =  \{11,02,20\} &\orez(2) &=0
\end{align*}
The induced quantum structure $1 \rrel{\eta} 3\times 3$ now relates $0$ with $00, 12$ and $21$, whereas the standard one relates $0$ with $00$, $11$ and $22$. Soon we shall see how this comes about.

\subsection{Representations of classical structures}\label{Every}
By definition \ref{frobdef}, classical structures are given as {\em internal\/} algebras. They are thus defined in any dagger monoidal category $\CCC$. However, some parts of the analysis of classical structures is simper with a more concrete representation. 

According to proposition \ref{unique}, a classical structure $(X,\mnd,\cmn,\unt,\cun)$ is completely determined by the monoid part $(X,\mnd,\unt)$. This internal monoid can be represented as a monoid of endomorphisms on $X$ in $\CCC$, as follows: proceeding as follows: 
\begin{itemize}
\item first {\em externalize\/} the internal monoid $(X,\nabla,\bot)$ in $\CCC$ as an ordinary monoid of vectors $\left(\CCC(X),\cdot,\bot\right)$, by setting
\bear
\varphi \cdot \psi & = & \nabla\circ (\varphi\otimes \psi)
\eear
\item then represent every vector $\varphi\in \CCC(X)$ as an action $\Upsilon\varphi$ over $X$, by
\bear
\Upsilon\ :\  \CCC(X) & \to &  \CCC(X,X)\\
\left(I\stackrel{\varphi}{\rightarrow} X\right) &\mapsto & \left(X\stackrel{\varphi\otimes X}\longrightarrow X\otimes X \stackrel{\nabla}\rightarrow X\right)
\eear
\end{itemize}
This second step can be viewed either as a generalization of Cayley's group representation to monoids, or as a special case of Yoneda's embedding of categories. 
\be{prop}
The monoid $\left(\CCC(X),\cdot,\bot\right)$ is isomorphic with the submonoid of $\left(\CCC(X,X),\circ,{\id}\right)$ which consists of the endomorphisms $f:X\rightarrow X$ such that $f\circ(a\cdot x)  =  (f\circ a)\cdot x$ holds for all $a,x\in \CCC(X)$. \ee{prop}
This allows representing any monoid as a monoid of endomorphisms. But those monoids that come from  a classical structure carry more. As observed in \cite{PavlovicD:MSCS08}, and further explored in \cite{VicaryJ:star}, the externalisation of every Frobenius algebra, and hence every classical structure, is also a $\starr$-algebra.  The categorical presentations \cite{JoyalA:Tannaka,VicaryJ:star} of the antilinear operation $\starr$ involve the formal duals, as spelled out in sec.~\ref{Frobenius}. 

\begin{defn}
An internal {\em $\starr$-monoid} in a monoidal category $\CCC$ is a structure $(X,X^\ast,\nabla,\bot,\starr)$ where 
\begin{itemize}
\item $(X,\nabla,\bot)$ is internal monoid
\item $X^\ast$ is a dual of $X$, and
\item $\starr : X \cong X^\ast$ is an isomorphism (always unitary). 
\end{itemize}
We write $\varphi^\starr = \starr\circ\varphi \in \CCC(X^\ast)$ for $\varphi\in \CCC(X)$.

A $\starr$-monoid homomorphism $f:X\to Y$ is a monoid homomorphism which moreover preserves the $\starr$, in the sense that
\[\xymatrix{
X^\ast \ar[r]^{f_\ast} \ar[d]_{\starr}& Y^\ast \ar[d]^{\starr}\\
X \ar[r]_f & Y
\\
}\]
commutes.
\end{defn}

\be{prop}
The monoid $\left(\CCC(X),\cdot,\bot\right)$ induced by a classical structure $(X,\nabla,\bot)$ comes with an involution
\bear
(-)^\starr\ :\  \CCC(X) & \to & \CCC(X)\\
\left(I\stackrel{\varphi}{\rightarrow} X\right) &\mapsto & \left(I\stackrel{\eta}\rightarrow X\otimes X \stackrel{\varphi^\ddag \otimes X}\longrightarrow X\right)
\eear
This involution is preserved by the representation $\Upsilon:\CCC(X) \to \CCC(X,X)$, in the sense that $\Upsilon\left(\varphi^\starr\right) = \left(\Upsilon \varphi \right)^\ddag$ holds.
\ee{prop}

\section{Simple classical structures in $\Rel$}\label{Simple}
In the rest of the paper, we explore and characterize classical structure in a category $\Rel$ of sets and relations. Any of its formalizations (some mentioned in the Introduction) will do. Computationally, relations are usually viewed as denotations of nondeterministic programs: a binary relation $R:A\rel B$ is the input/output relation of a program, which may output $b$ when given an input $a$ whenever $aRb$ holds \cite{MoggiE:notions}.

\subsection{Meaning of {\sf (spe)} in $\Rel$}
On the other hand, the isometry condition ({\sf spe}) here means that the relation $\mnd: X\times X\rel X$ is single-valued and surjective on $X$, i.e.
\begin{gather*}
\forall x,y,u,v\in X.\ x,y \in \mnd(u,v)\ \Longrightarrow\ x = y\\
\forall x\in X\exists uv\in X.\ x = \mnd(u,v)
\end{gather*}
Equivalently, ({\sf spe}) means that $\cmn =\mnd^{op}: X\rel X\times X$ injects $X$ into parts of $X\times X$ and is total on $X$.
\begin{gather*}
\forall x,y\in X.\ \cmn(x)\cap \cmn(y)\neq0\ \Longrightarrow\ x = y\\
\forall x\in X.\ \cmn(x) \neq 0 
\end{gather*}

\subsection{Meaning of {\sf (fro)} in $\Rel$ }
In the monoidal category $(\Rel,\times,1)$, the monoid action $\mnd:X\times X\rel X$ is a relation,  which assigns to every pair $x,y\in X$ a set $\mnd(x,y)\subseteq X$. The Frobenius condition ({\sf fro}) becomes
\bear
\left\{<x,y>\ |\ \mnd(i,j)\cap \mnd(x,y) \neq 0\right\} & =&  \left\{\left<x,\mnd(y',j)\right>\ |\ i\in \mnd(x,y') \right\} \\
& =&  \left\{\left<\mnd(i,x'),y\right>\ |\ j\in \mnd(x',y)
\right\}
\eear
This must be satisfied for all $i,j\in X$. 

\subsection{Meaning of $\sf (fro)\wedge (spe)$ in $\Rel$}\label{both}

\paragraph{Notation.} Since $\mnd(u,v)$, according to {\sf (spe)}, has at most one element, $\mnd$ is a partial operation. It is thus convenient to write it in the infix form whenever it is defined, i.e. $u\mnd v = \mnd(u,v) \neq 0$. 

The condition $\mnd(i, j)\cap \mnd(x,y) \neq 0$ now becomes $i\mnd j= x\mnd y$ and $i\in x\mnd y$ means that $i=x\mnd y$. The Frobenius condition thus boils down to
\begin{eqnarray}\label{group}
\{<x,y>\ |\ i\mnd j = x\mnd y\} & =&  \{<x,y'\mnd j>\ |\ i = x\mnd y' \} \\ & =&  \{<i\mnd x',y>\ |\ j = x'\mnd y\} \notag
\end{eqnarray}

This characterisation provides a rich source of classical structures.

\begin{prop}\label{abelianclas}
Every abelian group $(X,\mnd,\unt)$ in $\Set$ induces a classical structure in $\Rel$. 
\end{prop}

\bpr
If $(X,\mnd,\unt)$ is a group, then (\ref{group}) is satisfied by $x' = j\mnd y^{-1}$ and $y' = x^{-1}\mnd i$. 
\epr

\paragraph{Remark.} The nonstandard classical structures described in section \ref{Examples1} are easily seen to arise from the groups $\ZZz_2$ and $\ZZz_3$. 

\subsection{Simplicity}\label{Simplicity}
\be{defn}
A classical structure induced by an abelian group is called {\em simple}.
\ee{defn}

\be{prop}\label{nobase}
A simple classical structure, viewed as a comonoid, has only trivial subobjects. More precisely, any simple classical structure $Y$ is the range of exactly two comonoid monomorphisms: 
 \begin{itemize}
\item the empty relation $O:\emptyset \rel Y$ and 
\item the chaotic relation $I: 1 \rel Y$, where $0\in 1$ is related to every $y\in Y$.
\end{itemize}
\ee{prop}

\bpr
The fact that the empty relation $O:\emptyset \rel Y$ is a comonoid monomorphism is easily checked. We prove that the only nonempty subobject of a simple classical structure $Y$ is the chaotic relation $I: 1 \rel Y$.

A relation $R:X\rel Y$ is a monomorphism in $\Rel$ if and only if the induced map 
\bear
\wp R\ :\ \wp X & \to &  \wp Y\\
U & \longmapsto & \{y\in Y\ |\ \exists x\in U.\ xRy\} 
\eear
is injective. This implies 
\beq\label{nonempty}
\forall x\in X\exists y\in Y. \ xRy
\eeq
or else $\wp R\{x\} = \wp R\emptyset$.

On the other hand, given the comonoids $X$ and $Y$ in $\Rel$, unfolding the statement that a relation $R:X\rel Y$ is a comonoid homomorphism says that
\bea\label{cnd}
x\cmn <s,t>\ \wedge\ y\cmn <u,v> & \Longrightarrow & (xRy \iff sRu \wedge tRv)
\eea 
The claim is now that \eqref{nonempty} and \eqref{cnd} imply
\beq\label{alll}
\forall x\in X\forall u\in Y. \ xRu
\eeq
This claim is proven by instantiating \eqref{cnd} to $s = x$, $t = \bot$, with an arbitrary $u$, and $v=y-v$. The left-hand side of \eqref{cnd} is then satisfied, since $x\cmn <x,\bot>$ holds by definition, and $y\cmn<u,v>$ means $y = u+v$, for any simple $Y$. Together with \eqref{nonempty}, this instance of \eqref{cnd} implies $xRu$, i.e. \eqref{alll}. 

Thus $\wp R\{x\} = Y$ holds for all $x\in X$. Since $\wp R$ is injective, this implies that $X$ has at most one element, to which all of $Y$ is related. The relation $R$ is thus chaotic, as claimed.
\epr

\section{Classification of classical structures in $\Rel$}\label{All}

\subsection{$\starr$-algebras in $\Rel$}
In section \ref{Every}, we saw that every classical structure induces a $\star$-algebra. In $\Rel$, this restricts them to a very small family. The decomposition of Frobenius algebras into simple subalgebras follows.

\begin{prop}\label{partbij}
The representation $\repr : X \to \Rel(X,X)$ maps the elements of any classical structure $X$ in $\Rel$ to partial bijections.
\end{prop}

\bpr
We saw in section \ref{Simple} that classical structures in $\Rel$ are partial monoids, in the sense that $x\mnd y$ has at most one element. This means that for every $y\in X$ $\repr y: X\rel X$ is a partial map.

Since $\repr : X \to \Rel(X,X)$ is a $\starr$-representation, $\repr (y^\ddag) = \left(\repr y\right)^{op}$ is also a partial map. But a relation $R\in \Rel(X,X)$ such that both $R$ and $R^{op}$ are partial maps
\bear
xRy \wedge xRy' &\Longrightarrow & y = y'\\
xRy \wedge x'Ry & \Longrightarrow & x = x'
\eear
must be a partial bijection. In words, for every $x$ there is at most one $y$ such that $xRy$; and for every $y$ there is at most one $x$ such that $xRy$.
\epr

\subsection{The main results}
\begin{prop}\label{decomposition}
Every special Frobenius algebra in $\Rel$ is a biproduct of special Frobenius algebras where the unit is a singleton.
\end{prop}

\paragraph{Terminology.} The biproduct of sets $A$ and $B$ in $\Rel$ is simply their disjoint union $A+B$. This means that it is at the same time their product, and their coproduct. 

\bprf{ of proposition \ref{decomposition}}
Let the unit $\unt \in \Rel(X)$ of the special Frobenius algebra $X$ be $\unt = \{\phi_j\}_{j\in J}$. We claim that there is a partition
\bear
X & = & \bigcup_{j\in J} X_j
\eear
such that 
\bear
X_k \cap X_\ell \neq 0 &\Longrightarrow & k=\ell
\eear
and such that for every $j\in J$ the partial bijection $\repr \phi_j :X\rel X$ is just the identity on $X_j$. 

To prove this, note that 
\begin{itemize}
\item $\repr \unt = \id_X$,
\item $\repr \phi_j \subseteq \repr \unt$
\item if $a \in X_j\cap X_k$, then $a \repr \phi_j = a = a\repr \phi_k \Longrightarrow j=k$, because $a\repr : X\rel X$ must also be a partial bijection, as demonstrated in proposition \ref{partbij}.
\end{itemize}
Thus the domains of $\repr\phi_j$ must cover $X$, and they must be disjoint.

Now we claim that $(X_j,\mnd_j,\unt_j)$ is a submonoid of $(X,\mnd,\unt)$. This means that for every $x\in X_j$, the partial bijection $\repr x : X\to X$ restricts to a bijection $\repr_j x : X_j \to X_j$.

Suppose that for $x,y\in X_j$ it happens that $x\mnd y \in X_k$. That would mean that $y\mnd \phi_k$ must be defined, because $x\mnd y = (x\mnd y)\mnd \phi_k = x\mnd (y\mnd \phi_k)$. But then $y = y\mnd \phi_k \in X_k$, and we get $y\in X_j\cap X_k$. We have seen above that this implies $j = k$.
\epr

\begin{prop}\label{Frogroup}
Suppose that  $(X,\cmn,\mnd,\cun,\unt)$ is a classical structure in $\Rel$, such that the unit $\unt:1\to X$ is a function, i.e. a single element of $X$. Then the monoid $(X,\mnd,\unt)$ must be an abelian group in $\Set$. 
\end{prop}

\bpr
We first show that the monoid part of every classical structure $X$ in $\Rel$ must admit the inverses, as soon as it satisfies the assumptions of the proposition. More precisely, the claim is that condition (\ref{group}) from section \ref{both}, together with the assumption that the unit is a singleton  $\unt\in X$, implies that for every $k\in X$ there is $k^{-1}\in X$ such that $k\mnd k^{-1} = k^{-1}\mnd k = \unt$. 

First consider condition (\ref{group}) for $i = k$ and $j =\unt$. For the pair $<x,y> = <\unt,k>$, the second equation gives $x'\in X$ such that $\unt = x' \mnd k$. Dually, (\ref{group}) also holds for $i =\unt$ and $j=k$. For the pair $<x,y> = <k,\unt>$, the first equation gives $y'\in X$ such that $\unt = k\mnd y'$. Since $x' = x'\mnd \unt = x'\mnd (k\mnd y') = (x'\mnd k)\mnd y' = \unt \mnd y' = y'$, we can set $k^{-1} = x' = y'$.

To see that $(X,\mnd,\unt)$ is a group, it remains to be shown that the operation $\mnd$ is total, i.e. that $k\mnd \ell$ is defined for all $k,\ell\in X$. To see that this is the case, note that in each of the equations $\ell = (k^{-1}\mnd k) \mnd \ell = k^{-1} \mnd (k\mnd \ell)$, the left-hand side is defined if and only if the right-hand side is defined. Hence $k\mnd \ell$ must be defined.

Since the monoid operation $\mnd:X\times X \to X$ is total, and every element $k\in X$ has an inverse $k^{-1}$, we conclude that $(X,\mnd,\unt)$ is indeed a group.
\epr

Given an arbitrary classical structure $(X,\cmn,\mnd,\cun,\unt)$ in $\Rel$, we can now first apply proposition \ref{decomposition} to decompose it as a biproduct
\bear
X & = & \sum_{j\in J} X_j
\eear
of classical structures $(X_j,\cmn_j,\mnd_j,\cun_j,\unt_j)$ where each $\unt_j$ is a singleton. By proposition \ref{Frogroup}, each of these classical structures is a group. Hence the final result:

\begin{thm}
Every special Frobenius algebra in $\Rel$ is a biproduct (disjoint union) of groups. Every classical structure in $\Rel$ is a biproduct of abelian groups.
\end{thm}

Using this result, we can now effectively enumerate all classical structures in $\Rel$ with a given number of elements.

\subsection{Examples of classical structures}
Any classical structure over a set of $n$ elements can thus be constructed by choosing
\begin{itemize}
\item a partition $n = \sum_{j} n_j$, where $j\geq 1$,
\item an abelian group $X_j$ of order $n_j$ for each $n_j$.
\end{itemize}
For $n=2$, there are just two partitions: $n = 1+1$ and $n=2$. Since there is just one group with a single element, and just one group with 2 elements, these two partitions each determine a unique classical structure. They were described in section \ref{Examples1}.

For $n=3$, besides $n=1+1+1$ and $n=3$, we can also write $n=1+2$. The first two partitions give the classical structures described in section \ref{Examples1}. The nonstandard one comes from $\ZZz_3$. The third classical structure is the disjoint union $\ZZz_1 + \ZZz_2$.

For $n= 4$ there are five partitions. It is easy to see the pattern: e.g., $n= 2+2$ induces the classical structure $\ZZz_2 + \ZZz_2$, whereas $n=1+3$ induces $\ZZz_1+\ZZz_3$. Since there are two groups with 4 elements, $\ZZz_4$ and the Kleinian group $D_4 = \ZZz_2\times \ZZz_2$, the trivial partition $n=4$ induces two different classical structures. They both have the same classical element, consisting of all of $n=4$; but they induce different quantum structures, entangling each element with its group inverse. Since each element of $D_4$ is its own inverse, its quantum structure coincides with the one induced by the standard classical structure $\ZZz_1+\ZZz_1+\ZZz_1+\ZZz_1$. In any case, there are exactly 6 different classical structures with 4 elements.

For $n=5$, there are 7 different partitions, and they induce 8 different classical structures. E.g., the partition $n = 2+3$ corresponds to the classical structure $\ZZz_2+\ZZz_3$, with the classical elements $\{0,1\}$ and $\{2,3,4\}$. The quantum structure is $\eta = \{00,11,22,34,43\}$. 

For $n=6$, there are 11 partitions. There are 2 groups with 6 elements, but only the cyclic one is abelian\ldots

In all cases, the classical elements of a classical structure are just the underlying sets of its constituent groups. They do not depend on the actual structure of the groups. This structure is, however, reflected in the induced quantum structure, which entangles each element with its group inverse.

\section{Conclusions and future work}\label{Conclusions}
We classified classical structures in $\Rel$, and found that many are nonstandard. They also induce many nonstandard quantum structures $I\tto{\eta}X\times X$ in $\Rel$. If $X$ is a group, then $\eta$ entangles each $a\in X$ with its inverse $a^{-1}$. Moreover, each of the nonstandard classical structures induces a nonstandard abstraction operator $\kappa x$, binding the variable $x$ in the polynomial relations $\varphi(x)\in \Rel[x]$. For monoidal categories in general, such operations and their meaning were analyzed in \cite{PavlovicD:MSCS97}. 
In $\Rel$ in particular, the situation seems rather curious. While the nonstandard classical structures support specifying relational polynomials, as nondeterministic programs with nonstandard variables --- proposition \ref{nobase} says that there are only trivial classical elements to be substituted for these variables. The distinctions of the elements belonging to the same group within a nonstandard classical structure turn out to be classically indistinguishable. However, this indistinguishability, observed through a different classical structure, can be used as a computational resource. Indeed, switching between the different classical structures in order to use this resource is the essence of some of the most important quantum algorithms. The interesting structural repercussions of this method within the relational view of nondeterministic computation need to be further explored in future work.

\paragraph{Acknowledgements.} I am grateful to Ross Duncan and Chris Heunen for questions and comments.

\bibliographystyle{plain}
\bibliography{ref-epistemic,../PavlovicD}

\end{document}